%% file: main.tex
\pdfoutput=1
\documentclass[%
 reprint,
nofootinbib,
 amsmath,amssymb,
 aps,
]{revtex4-2}
\usepackage{amsmath}
\usepackage{tikz}
\usetikzlibrary{quantikz}
\usepackage[caption=false]{subfig}
\usepackage{array}
\usepackage{dsfont}

\usepackage{multirow}

\usepackage{graphicx}
\usepackage{dcolumn}
\usepackage{bm}
\usepackage{subfiles}
\usepackage[normalem]{ulem}

\usepackage{natbib}
\usepackage{hyperref}
\usepackage{titlesec}

\titleformat{\paragraph}
{\normalfont\normalsize\bfseries}{\theparagraph}{1em}{}
\titlespacing*{\paragraph}
{0pt}{3.25ex plus 1ex minus .2ex}{1.5ex plus .2ex}

\begin{document}

\preprint{APS/123-QED}


\title{Combining Matrix Product States and Noisy Quantum Computers for Quantum Simulation}
\author{Baptiste Anselme Martin$^{1,2}$}%
\author{Thomas Ayral$^{3}$}%
\author{François Jamet$^{4}$}%
\author{Marko J. Ran\v{c}i\'{c}$^1$}%
\author{Pascal Simon$^2$}
 \affiliation{1. 
 TotalEnergies, Tour Coupole - 2 place Jean Millier 92078 Paris La D\'{e}fense, France}
 \affiliation{2. Universit\'{e} Paris-Saclay, CNRS, Laboratoire de Physique des Solides, 91405, Orsay, France}
 \affiliation{3. Eviden Quantum Laboratory, Les Clayes-sous-Bois, France}
 \affiliation{4. National Physical Laboratory, Teddington, TW11 0LW, United Kingdom}

\date{\today}

\begin{abstract}
	
Matrix Product States (MPS) and Operators (MPO) have been proven to be a powerful tool to study quantum many-body systems but are restricted to moderately entangled states as the number of parameters scales exponentially with the entanglement entropy.  While MPS can efficiently find ground states of 1D systems, their capacities are limited when simulating their dynamics, where the entanglement can increase ballistically with time.  On the other hand, quantum devices appear as a natural platform to encode and perform the time evolution of correlated many-body states. However, accessing the regime of long-time dynamics is hampered by quantum noise. In this study we use the best of worlds: the short-time dynamics is efficiently performed by MPSs, compiled into short-depth quantum circuits, and is performed further in time on a quantum computer thanks to efficient MPO-optimized quantum circuits. We quantify the capacities of this hybrid classical-quantum scheme in terms of fidelities  taking into account a noise model. We show that using classical knowledge in the form of tensor networks provides a way to better use limited quantum resources and lowers drastically the noise requirements to reach a practical quantum advantage. Finally we successfully demonstrate our approach with an experimental realization of the technique. Combined with efficient circuit transpilation we simulate a 10-qubit system on an actual quantum device over a longer time scale than low-bond-dimension MPSs and purely quantum Trotter evolution.

\end{abstract}
\maketitle

\section{Introduction}

Quantum computers hold the promise of solving problems intractable for their classical counterparts \cite{feynman2018simulating, universalqsim}. With the advent of the Noisy Intermediate Scale Quantum (NISQ) devices \cite{preskill2018quantum}, the study of quantum many-body systems is believed to be one of the first applications of quantum computers. By manipulating superposed and entangled qubit states, quantum devices appear as a natural platform to encode correlated many-body states. Numerous quantum algorithms have been proposed to solve quantum many-body systems with applications for condensed-matter physics, electronic structure problems, or quantum chemistry. Quantum Phase Estimation (QPE) \cite{kitaev1995quantum} or Variational Quantum Eigensolver (VQE) \cite{peruzzo2014variational} are the most prominent algorithms proposed to determine the ground state and the energy of a many-body Hamiltonian $\mathcal{H}$. Alongside the search for ground states, the simulation of quantum dynamics is another key challenge, where states can be brought far from equilibrium with a high level of entanglement. Algorithms have been proposed and tested to perform real-time simulation of many-body systems on a quantum computer \cite{universalqsim, miessen_quantum_2022, li_efficient_2017, yuan_theory_2019, smith_simulating_2019, cirstoiu_variational_2020, benedetti_hardware-efficient_2021, barison_efficient_2021,barratt_parallel_2021}.  However, current devices still suffer from significant levels of noise, which strongly limit their capacities. 

On the other hand, numerical computational techniques have been increasingly successful to study quantum systems. Among them, tensor networks provide an efficient way to represent correlated quantum states \cite{Schollw_ck_2011, Or_s_2014, RevModPhys.93.045003}. In one dimension, Matrix Product States (MPS) allow us to find ground states of local gapped Hamiltonian thanks to the famous density matrix renormalization group (DMRG) \cite{PhysRevLett.69.2863} algorithm or to simulate their dynamics with, for instance, the Time Evolution Block Decimation (TEBD) algorithm \cite{suzuki_generalized_1976}. Moreover, tensor network techniques are competitive to simulate quantum computers in the presence of noise \cite{zhou_what_2020, ayral_density-matrix_2022, noh_efficient_2020, zhang_entanglement_2022}. More recently, the interplay between quantum simulations and tensor networks was highlighted by IBM's kicked-Ising simulation on a 127-qubit device \cite{Kim2023}, where tensor networks proved to be powerful tools to verify the experiment's results in nontrivial regimes and provided valuable insights on the computational hardness of such quantum simulations \cite{tindall_efficient_2023, begusic_fast_2023, liao_simulation_2023, patra_efficient_2023}. This indicates that quantum computer simulations with naive approaches are not able today to outperform tensor network simulations. On the other hand, physically relevant problems still remain out of reach for state-of-the-art tensor network techniques, because they fail to represent highly entangled states. A prototypical example is the dynamics of global quenched quantum systems, where an initial state is abruptly driven by a Hamiltonian. Such systems typically exhibit a ballistic growth of the entanglement with time \cite{Alba_2018}, implying that tensor network techniques can only access short-time dynamics.

\begin{figure}[!t]
	\centering
	\includegraphics[width = 0.95\linewidth]{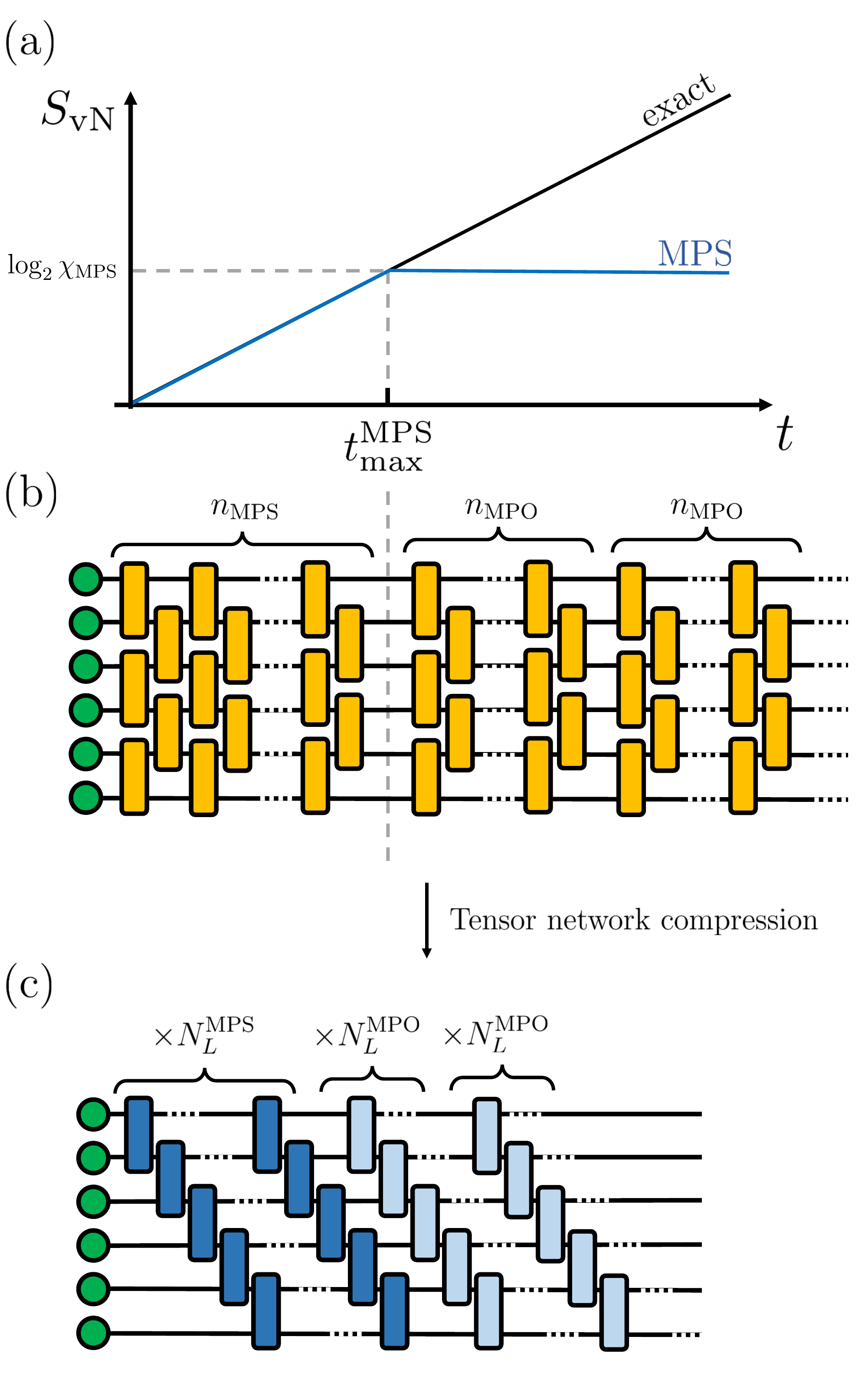}
	\caption{(a) Classical MPS of dimension $\chi_\text{MPS}$ can only access short-time dynamics. The maximum simulated time $t_\text{max}^\text{MPS} $ is defined as the time where MPS's entanglement entropy $S_\text{vN}$ reaches $\log_2{\chi_\text{MPS}}$. (b) The quantum circuit corresponding to the TEBD algorithm. The time-evolved state at time $t \leq t_\text{max}^\text{MPS} $ can be classically represented by a MPS, and can be efficiently represented by a circuit (c) of depth $N_L^\text{MPS}$. The time-evolution can be carried further on the quantum computer, thanks to quantum circuits of depth $N_L^\text{MPO}$ optimized by MPOs.}
	\label{fig:schemeQMPSO}
\end{figure}

In this paper, we bridge the gap between quantum computer and tensor network simulations. We utilize the combination of MPS solutions that are tractable on a classical computer and quantum circuits on noisy quantum devices to study the dynamics of a paradigmatic spin chain Hamiltonian. This work is organized as follows. After introducing the key concepts of MPS and their time evolution in Sec.\ref{sec:MPS}, we detail the method used to optimize quantum circuits thanks to MPSs and MPOs (for Matrix Product Operators, operators encoded in MPSs \cite{PhysRevLett.93.207204, Pirvu_2010}) in Sec.\ref{sec:qmps}. In Sec. \ref{sec:results}, we investigate the time evolution of a global quench in the one-dimensional transverse field Ising model where short-time simulations are performed by MPS techniques, leading to optimized quantum circuits that are extended for longer time simulations on a noisy quantum computer. To quantify our results, we use the fidelity as a performance metric in Sec. \ref{subsec:fid}. Finally in Sec. \ref{sec:exp}, elaborate noise mitigation techniques allow us to test our scheme on actual quantum devices from IBM Quantum \cite{IBM}. This hybrid classical-quantum procedure is summarized in Fig. \ref{fig:schemeQMPSO}.

\section{Matrix Product States and their Time Evolution}\label{sec:MPS}

In this work, we employ numerical simulations based on MPS, which are a class of one-dimensional many-body quantum states that allow an efficient representation of entanglement. Let us first briefly introduce for completeness the key concepts around MPS.
We consider $N$ spins, each spin being described by a local Hilbert space $\mathcal{H}_{loc}$ of dimension $d$. Let $\{\ket{\sigma}\}$ be an orthonormal basis of $\mathcal{H}_{loc}$. A general many-body state $ \ket{\Psi}$ living in $\mathcal{H} = \bigotimes_{i=1}^{N} \mathcal{H}_{loc}$ can be written as:
\begin{equation}
    \ket{\Psi} = \sum_{\sigma_1, ..., \sigma_N} c_{\sigma_1 ... \sigma_N} \ket{\sigma_1 ... \sigma_N},
\end{equation}
where $c_{\sigma_1 ... \sigma_N} \in \mathbb{C}$ and $\ket{\sigma_1 ... \sigma_N} = \ket{\sigma_1}\otimes ... \otimes \ket{\sigma_N}$.
A MPS is a class of quantum many-body states where the coefficients $c_{\sigma_1 ... \sigma_N}$ are obtained by multiplying matrices $\{A_n^{\sigma_n}\}$ together. For an open boundary state, a MPS reads as:
\begin{align}
    \ket{\Psi}  
    &= \sum_{\{\sigma_n\}} \sum_{\{\alpha_n\}} A^{[1]\sigma_1}_{\alpha_1} A^{[2]\sigma_2}_{\alpha_1\alpha_2}  ... A^{[N]\sigma_N}_{\alpha_{N-1}} \ket{\sigma_1 \sigma_2... \sigma_N},
\end{align}
where $A^{[n]\sigma_n}$ for $2\leq n \leq N-1$  are complex matrices of dimension $\chi_{n-1}\times\chi_{n}$, where $\chi_n$ is called the bond dimension between spin sites $n-1$ and $n$. At the boundaries, the tensors $A^{[1]\sigma_1}$ and $A^{[N]\sigma_N}$ are vectors of dimension $\chi_1$ and $\chi_{N-1}$. The bond dimension of a MPS dictates the storage cost of the state and is a key element to understand how entangled the MPS is. Indeed, by considering the bipartition of the state of the Hibert space as $\mathcal{H} = \mathcal{H}_R\otimes \mathcal{H}_L$, where $\mathcal{H}_R$ ($\mathcal{H}_L$) denotes the Hilbert space of the subsystem at the left (right) of the bond between the spin sites $n$ and $n+1$, we can rewrite the state $\ket{\Psi}$ as
\begin{equation}\label{eq:schmidt}
    \ket{\Psi} =\sum_{k=1}^{\chi_{n}} \lambda_k \ket{\Psi^L_k}\ket{\Psi^R_k},
\end{equation}
where $\ket{\Psi^{L(R)}_k} \in \mathcal{H}_{L(R)}$ and $\lambda_k \in \mathds{C}$. Eq. (\ref{eq:schmidt}) is called the Schmidt decomposition of $\ket{\Psi}$. In this form, it becomes possible to quantify the amount of entanglement betwen the subsystem $L$ and $R$ by calculating the entanglement or von Neumann entropy. Starting from the density matrix $\rho = |\Psi \rangle \langle \Psi |$, we define the reduced density matrices $\rho_{R(S)} = \text{Tr}_{S(R)} \rho$. The von Neumann entropy is then calculated as
\begin{align}
       S_{\text{vN}} &= -\text{Tr}_R\Large[ \normalsize\rho_R \text{log}_2(\rho_R) \Large]\\
      &= -\text{Tr}_L\Large[  \normalsize \rho_L \text{log}_2(\rho_L) \Large],
\end{align}
Using the Schmidt decomposition of $\ket{\Psi}$ as in Eq. (\ref{eq:schmidt}), we directly obtain $S_{\text{vN}}$ as
 \begin{equation}
     S_{\text{vN}} = -\sum_{k=1}^{\chi_n} \lambda_k^2\text{log}_2 \lambda_k^2.
 \end{equation}
The entanglement entropy is maximal when all the Schmidt values $\{\lambda_k\}$ are all equal, {\it i.e.} 
$\forall k$, $\lambda_k = 1/\sqrt{\chi_n}$ (as $\ket{\Psi}$ is normalized). Therefore the maximal amount of entanglement entropy between $L$ and $R$ is equal to $\log_2 \chi_n$.  This highlights that the amount of entanglement in a MPS is directly related to the dimensions of its tensors and therefore to its storage cost.

In this study, we aim at benchmarking the capabilities of quantum computers compared to standard MPS techniques for time evolution \cite{Paeckel_2019}. To do so, we study the simulation of global quantum quenches where we consider an initial state $\ket{\Psi_0}$ that is suddenly driven under a Hamiltonian at $t\textgreater 0$. The time-evolved state reads as
\begin{equation}
    \ket{\Psi(t)} = e^{-iHt}\ket{\Psi_0}.
\end{equation}
As an example, we consider here the paradigmatic one-dimensional spin-1/2 Ising model with a transverse field. The corresponding Hamiltonian is defined as
\begin{equation}
 H = -J\sum_{i=1}^{L-1} X_i X_{i+1} - h\sum_{i=1}^{ L}Z_i 
\end{equation}
where $L$ is the number of spin sites on the chain, $J$ is the interaction strength between neighboring spins and $h$ is the transverse field value. The spin operators $X_i$ and  $Z_i$  are equals to $\begin{pmatrix} 0&1\\1&0\end{pmatrix}$ and  $\begin{pmatrix} 1&0\\0&-1\end{pmatrix}$ respectively.
\begin{figure}
	\centering
	\includegraphics[width=0.9\linewidth]{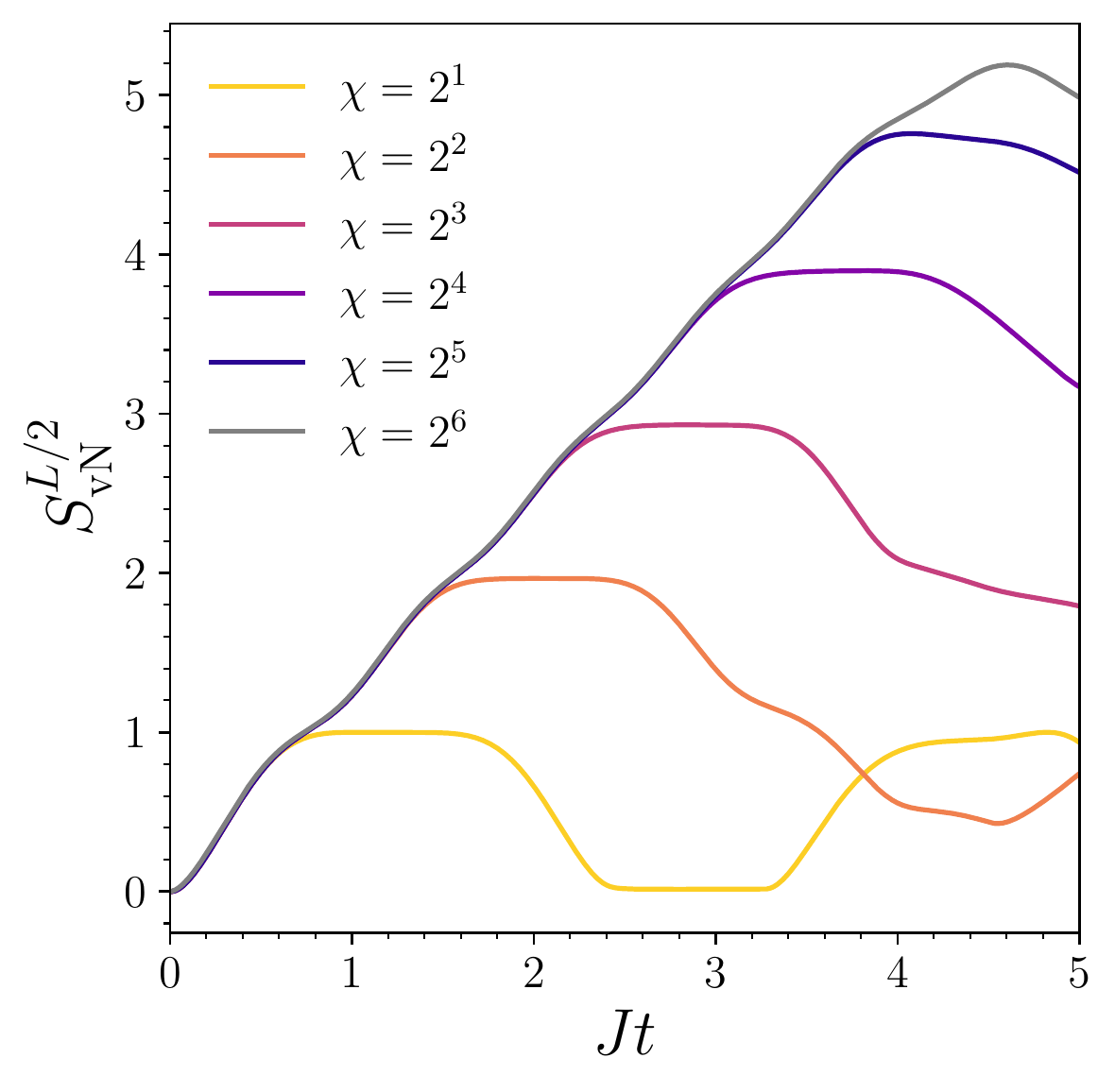}
	\caption{The von Neumann entropy $S_{\text{vN}}$ at half-chain as a function of time from MPS states with different bond dimensions $\chi$ for $L=10$ spin sites under a global quench. $S_{\text{vN}}$ grows ballistically with time and saturates at $\log_2 \chi $. For 12 sites, the MPS of bond dimension $2^6$ is not truncated and therefore is exact.}
	\label{fig:entanglemententropyl10mpsdt01}
\end{figure}

The initial state is chosen to be a Néel state defined as $\ket{\uparrow\downarrow\uparrow\downarrow ...}$. This global quantum quench brings the state far from equilibrium, whose dynamics are characterized by a ballistic growth of the entanglement with time \cite{Alba_2018}. As a consequence, only short-time dynamics are accessible t MPS simulations, as they can handle only weakly entangled states. The simulations of such physical systems provide relevent and interesting benchmarking problems for quantum computers, as demonstrated on IBM's experiment of a transverse field Ising system on a 127-qubit quantum processor \cite{Kim2023}.

To simulate time-evolution with MPS, we use the so-called TEBD algorithm or Trotter Decomposition method \cite{suzuki_generalized_1976}. We first formulate the Hamiltonian $H= \sum_{i=1}^{N-1} h_{i,i+1}$ as a sum of local Hamiltonians
defined as
\begin{align}
    h_{i,i+1} &= -JX _iX _{i+1} -\frac{h}{2}( Z_i + Z_{i+1}) \text{     } \forall i \in [2,L-2],\\
    h_{1,2} &= -J X_1 X_{2} -h(Z_1 + \frac{1}{2}Z_2),\\
    h_{L-1,L} &= -JX_{L-1}X_{L}  -h(\frac{1}{2} Z_{L-1} + Z_{L}).\\
\end{align}
Then, we approximate the time propagator $U(t) = e^{-iHt}$ by performing time steps of duration $dt$ thanks to the standard first-order Trotter-Suzuki's formula as
\begin{equation}
    U(t) \simeq \prod_{n=1}^{t/dt} \prod_{i \text{ even}} e^{-ih_{i,i+1}dt} \prod_{j \text{ odd}} e^{-ih_{j,j+1}dt}.
\end{equation}
At each time step, we apply local unitary transformations $e^{-ih_{i,i+1}dt}$ between two neighboring sites, which leads to an update of the corresponding tensors. This leads to an increase in the bond dimensions as the state will get more entangled. However, when bounding the bond dimension to a maximal value $\chi_\text{max}$, we need to truncate the state by only keeping the $\chi_\text{max}$ largest Schmidt values of the specific bond. By doing so, we limit the entanglement entropy carried by the MPS. To illustrate this, Fig. \ref{fig:entanglemententropyl10mpsdt01} shows the entanglement entropy at the middle of the chain $S_\text{vN}^{L/2}$ as a function of time for MPS with different bond dimensions for the critical point $J/h = 1$. This parameter choice will be kept throughout the rest of this work. As the entanglement entropy grows with time, the MPS is not able to efficiently represent long-time-evolved states that carry more entanglement than it can represent.

\section{Optimizing Quantum Circuits with Tensor Networks}\label{sec:qmps}

\subsection{Quantum Matrix Product States (QMPS)}
\begin{figure*}[!t]
	\centering
	\includegraphics[width=1.\linewidth]{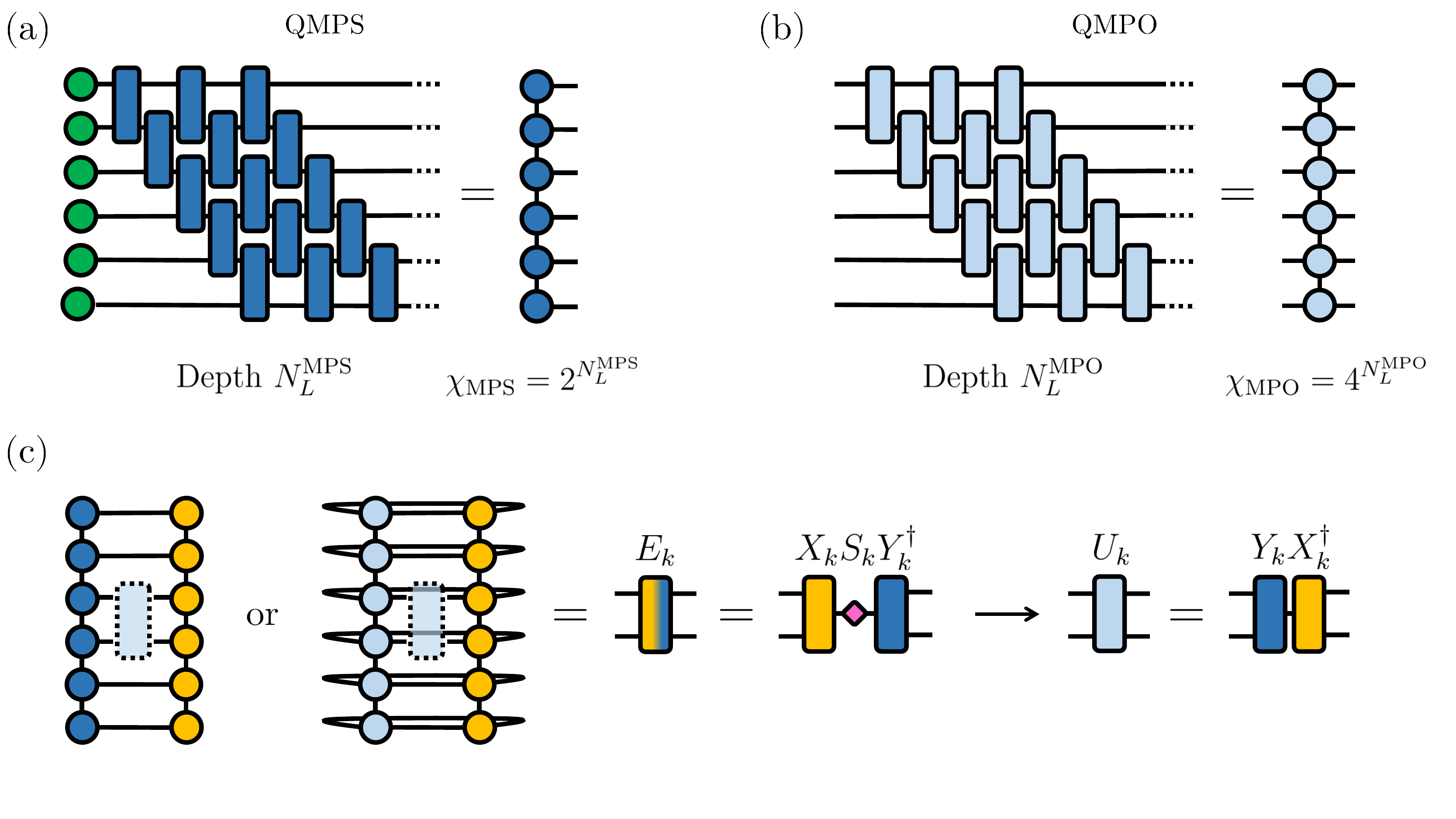}
	\caption{(a) A state generated by a sequential quantum circuit of depth $N_L^\text{MPS}$ can be exactly represented by a MPS of bond dimension $2^{N_L^\text{MPS}}$. (b) Similarly, a sequential quantum circuit, of depth $N_L^\text{MPO}$, without specifing any input state, can be exaclty represented by a MPO of bond dimension $4^{N_L^\text{MPO}}$. (c) To optimize the unitaries composing a QMPS or QMPO in order to maximize its overlap with a MPS or  MPO representation of a quantum circuit, we calculate the environment tensor of each unitary. The unitary is then updated from the environment tensor's SVD. }
	\label{fig:qmps_qmpo}
\end{figure*} 

In Section \ref{sec:MPS}, we detailed the MPS formalism, how it can be used to simulate the dynamics of quantum systems, but most importantly its bottleneck, lying in its ability to represent states with a high degree of entanglement. On the other hand, quantum computers do not suffer from these limitations in principle as they naturally use entanglement but they are limited by noise. Representing MPSs with efficient quantum circuits is an appealing challenge as it offers a promising pathway to extend classical methods beyond their limitations, such as performing time evolution or preparing highly entangled states, while optimizing the use of quantum resources.

The preparation of MPS states with qubits was first presented in \cite{schonsequential2005}. Later, techniques were developed to generate quantum circuits approximating a given MPS \cite{PhysRevA.101.010301, ran_encoding_2020,shirakawa_automatic_2021, dov2022approximate,rudolph_decomposition_2022}, or alternatively optimize quantum circuits using a MPS representation \cite{lin_real-_2021}. 
 Among potential applications, it has been proposed to initialize parametrized quantum circuits with classically optimized tensor network states \cite{dborin_matrix_2021, rudolph_synergy_2022, preentanglingcqc}. Tensor networks have also inspired classes of variational quantum circuits \cite{huggins_towards_2019, liu_variational_2019,miao_quantum-classical_2021, PhysRevLett.127.040501, barratt_parallel_2021, smith_crossing_2022, barratt_parallel_2021,haghshenas_variational_2022, astrakhantsev2023time}  and have been experimentally realized on quantum devices \cite{dborin_simulating_2022,chertkov_holographic_2022, foss-feig_entanglement_2022, niu_holographic_2022}. Embedding tensor networks into quantum algorithms has also been explored in the context of DMFT calculations \cite{jamet2023anderson}.

In this section, we use an approach similar to the algorithm defined in \cite{lin_real-_2021}. The goal here is to classically optimize a quantum circuit thanks to MPS techniques to prepare a state approximating the time-evolved state $\ket{\Psi(t)} = e^{-iHt}\ket{\Psi_0}$. To do so, we use a circuit ansatz built from staircase layers of two-qubit gates, as shown with Fig. \ref{fig:qmps_qmpo}a. With this specific ansatz, a quantum circuit of depth $N_L^\text{MPS}$ generates a state that can be exactly represented with a MPS of bond dimension $\chi_\text{MPS} = 2^{N_L^\text{MPS}}$. The resulting state will be referred to as ``Quantum Matrix Product State" (QMPS), and can be written as

\begin{equation}
    \ket{\Psi_{\text{QMPS}}(t)} = \prod_k U_k(t) \ket{\Psi_0},
\end{equation}
where $U_k(t)$ are unitary gates that compose the circuit acting on the initial state $\ket{\Psi_0}$. To approximate the state at time $t$, we classically optimize the unitaries $\{U_k(t) \}$ to maximize the overlap with $\ket{\Phi_{\text{MPS}}(t)}$, obtained from a TEBD simulation with a MPS of bond dimension $\chi_\text{MPS} = 2^{N_L^\text{MPS}}$. To do so, we iteratively optimize each unitary $U_k$ to maximize the overlap $F_k$ between the target state $\ket{\Phi_{\text{MPS}}}$ and $\ket{\Psi_{\text{QMPS}}} = \prod_k U_k \ket{\Psi_0}$ as
\begin{align}
    F_k &= \langle \Phi_{\text{MPS}} | (\prod_{i>k} U_i) U_k \prod_{j<k} U_j \ket{\Psi_0} \\
    &= \langle \Phi_{k+1} | U_k \ket{\Psi_{k-1} }\\
    &= \text{Tr}(E_k U_k),
\end{align}
where the environment tensor $E$ is defined as 
\begin{equation}
    E_k = \text{Tr}_{\overline{k}}(\ket{\Psi_{k-1}}\langle\Phi_{k+1}|).
\end{equation}
Here $\text{Tr}_{\overline{k}}(.)$ operates the trace over all qubits but the qubits acted on by the unitary $U_k$. Graphically, $E_k$ is calculated by contracting all qubit indices between the ones involved in the unitary operation $U_k$. By forming a Singular Value Decomposition (SVD) of $E_k$, we rewrite $E_k$ as
\begin{equation}\label{eq:Esvd}
    E_k = X_k S_k Y_k^\dagger .
\end{equation}
From Eq. (\ref{eq:Esvd}) the unitary $U_k$ is then updated according to \cite{evenbly2009algorithms} by
\begin{equation}
  U_k \leftarrow Y_k X_k^\dagger.
\end{equation}
\begin{figure}[h]
	\centering
	\includegraphics[width=0.9\linewidth]{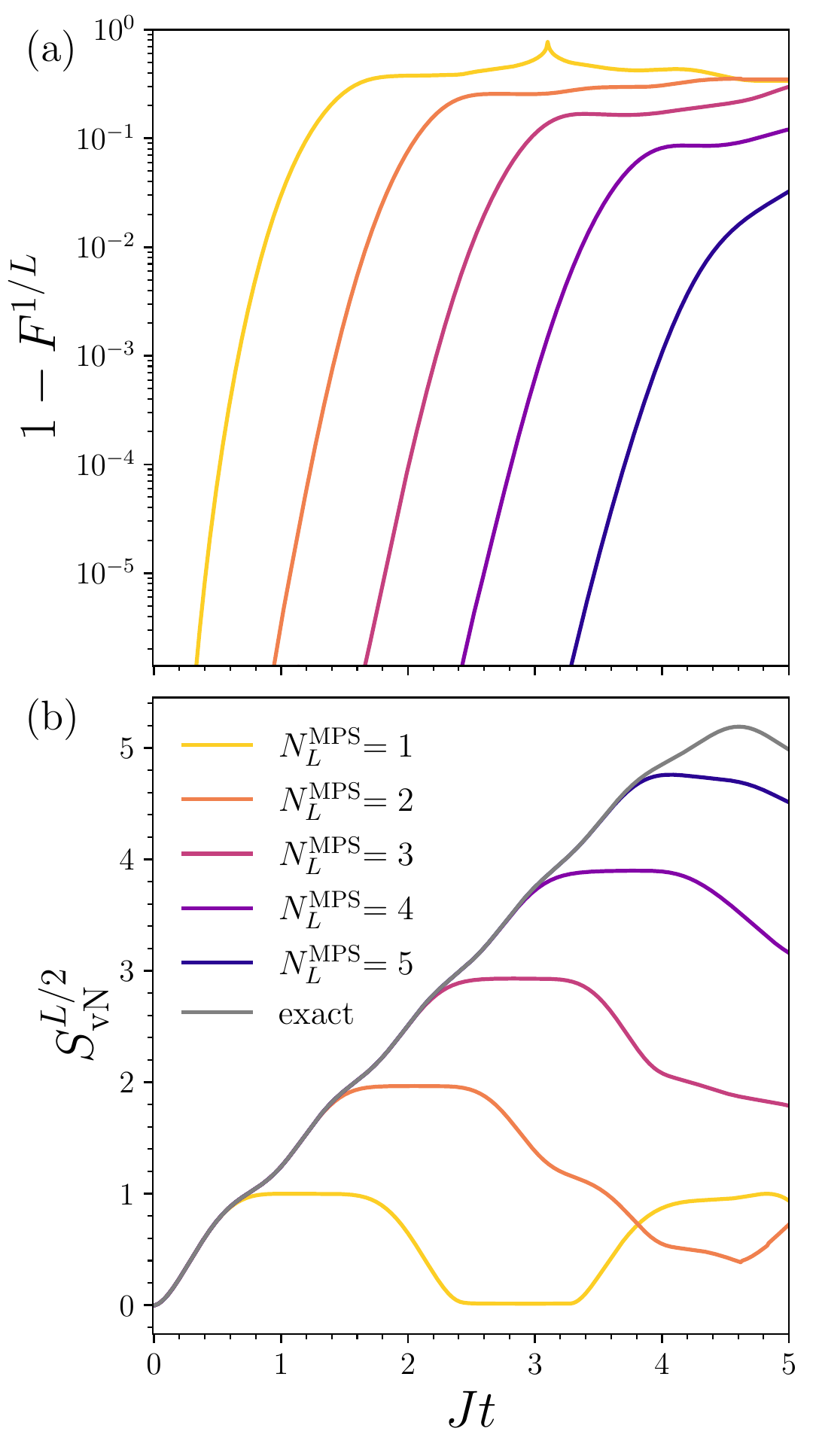}
	\caption{The infidelity per site (a) and the entanglement entropy (b) as a function of time for classically optimized quantum circuits of different number of layers $N_L^\text{MPS}$ using MPS of corresponding bond dimension $\chi = 2^{N_L^\text{MPS}}$, for $L = 12$ system.}
	\label{fig:F_S_QMPS}
\end{figure}

This optimization step is performed for each unitary $U_k$ consecutively. The procedure is sketched in Fig. \ref{fig:qmps_qmpo}c. The whole process is repeated until either a convergence criterion or a maximum number of sweeps is reached. In practice, to obtain the state $|\Psi_{\text{QMPS}}(t+dt)\rangle$, we optimize the overlap with $|\Phi_{\text{MPS}}(t+dt)\rangle$ using the unitaries $\{U_k(t)\}$ as a starting point.

In Fig. \ref{fig:F_S_QMPS}, we show the performance obtained from the optimized QMPS $|\Psi_{\text{QMPS}}(t)\rangle$.  We take exact time-evolved states with a time step $dt = 0.01$  (in $1/J$ units, as it will be implicitly assumed throughout the rest of the manuscript) as a reference solution  $|\Psi_\text{ref}(t)\rangle$. In Fig. \ref{fig:F_S_QMPS}a, we evaluate the quality of the QMPS states of different layers with the infidelity per site calculated as $1-F^{1/L}$, where $F = |\langle \Psi_\text{ref}(t)|\Psi_{\text{QMPS}}(t)\rangle|^2$. With Fig. \ref{fig:F_S_QMPS}b, we observe the same behavior as in Fig \ref{fig:entanglemententropyl10mpsdt01} in terms of entanglement entropy. As expected, at a fixed time $t$, the quality of the result increases when the depth (and the bond dimension) increases. Moreover, when $t$ increases, the state becomes more entangled and the fidelity decreases. Most importantly, both QMPS and MPS fail at the same simulation time, when the entanglement entropy reaches its maximum with respect to $N_L^\text{MPS}$  and $\chi_\text{MPS}$. In this example, QMPS states demonstrate similar strengths (and weaknesses) to MPSs while being implementable on a quantum computer. This method provides a powerful way to find entanglement-efficient circuits for time evolution that exploits a DMRG-like optimization, as opposed to the usual Trotter-Suzuki circuits that are particularly inefficient in terms of entanglement-per-added-gate.

\subsection{Quantum Matrix Product Operators (QMPO)}

In the previous section, we have shown that we can encode time-evolved MPSs as quantum circuits. We want now to reach longer simulation time by  time-evolving QMPSs on a quantum computer. Despite having reached the classical limit (set by the bond dimension) for the time-evolved state, it remains possible to further classically manipulate and optimize parts of the quantum circuit that will be append to the QMPS circuit. In this section, we propose to find better quantum circuits for the Trotter time evolution by leveraging Matrix Product Operators (MPO), a one-dimensional tensor network, similar to MPS, encoding operators acting on quantum states. 
Similarly to a MPS, a MPO is also characterized by its bond dimension, which relates the classical cost to store and manipulate the operator. To exactly represent a general $L$-qubit operator, we need a MPO of bond dimension $4^{L/2} = 2^L$. However, quantum circuits that generate a moderate amount of entanglement can be efficiently encoded as MPOs of moderate bond dimension. In this section, we compress Trotter quantum circuits into short-depth quantum circuits. We use the same approach as the one used for the optimization of QMPSs. First, as shown in Fig. \ref{fig:qmps_qmpo}a that sequential quantum circuit of depth $N_L^{\text{MPO}}$ can be represented exactly as a MPO of bond dimension $4^{N_L^{\text{MPO}}}$. These circuits will also be our ansatz to encode Trotter circuits. A MPO representation of the Trotter circuits can be obtained by performing the TEBD algorithm on the identity MPO defined as the tensor product of the single-qubit identity operators $\otimes_{i=1}^L \mathds{1}_i$,  whose MPO pictorial representation is simply $L$ parallel lines. Then, the gates are applied sequentially onto pairs of local tensors, which are then split by an SVD, in the same fashion as the MPS algorithm. By doing so for $n$ Trotter steps, we obtain the MPO approximation for the time evolution operator at time $ndt$. We fix the bond dimension to $4^{N_L^{\text{MPO}}}$, which limits the simulation time or circuit depth that we can efficiently encode as a MPO. Here we use again Trotter circuits with $dt = 0.01$ as our reference circuits.

To evaluate the fidelity $F_{\text{op}}$ between two unitaries $U$ and $V$ acting on $L$ qubits, we use the Frobenius norm defined as 
\begin{equation}
	F_{\text{op}} = \frac{\text{Tr}(U^\dagger V)}{2^L}.
\end{equation} 
Calculating $\text{Tr}(U^\dagger V)$ where two $U$ and $V$ are MPOs of bond dimension $\chi_\text{MPO}$ with local physical dimension $d$ ($=2$ in case of qubits) can be done by contracting the tensor network shown in Fig.\ref{fig:qmps_qmpo}c, whose computational cost scales as $\mathcal{O}(L(d^2\chi_\text{MPO})^3)$. It is important to note that if the MPS simulations and QMPS optimizations are performed with bond dimension $\chi_\text{MPS}$ (with a computational cost scaling as $\mathcal{O}(L(d\chi_\text{MPS}^3)$ ), the maximum MPO bond dimension $\chi_\text{MPO}$ must be set accordingly as $\chi_\text{MPS}/d$ ($\chi_\text{MPS}/2$ for qubits).  We also emphasize that exact MPS simulations can be done with a bond dimension $2^{L/2}$. Therefore, the maximum relevant MPO bond dimension is bounded by $2^{L/2}/2 =2^{L/2 -1}$, which leads to a maximum number of layers equal to $\lfloor\log_4 (2^{L/2 -1}) \rfloor$ ($\lfloor. \rfloor$ designates the floor function). 

In the same fashion as for QMPSs, we want to maximize the overlap between the quantum circuit ansatz from Fig.\ref{fig:qmps_qmpo}a (that we will call Quantum Matrix Product Operator or QMPO) and the MPO approximation of the time evolution operator $U_{\text{Tr}}$. To optimize a given unitary $U_k$ from the QMPO, we calculate the environment tensor of $U_k$ by removing the gate from the overlap tensor contraction. The gate $U_k$ is then updated from the SVD of the environment tensor,as shown in Fig.\ref{fig:qmps_qmpo}c. Note that similar approaches \cite{Tepaske_2023, Mc_Keever_2023} proposed to optimize quantum circuits for quantum simulation with MPOs by minimizing variational parameters of a given quantum circuit ansatz with an optimizer, while here the optimization is performed by a purely tensor network technique\footnote[1]{During the elaboration of this manuscript, we came to be aware of two related works. Ref. \cite{keever2023adiabatic} propose the use of MPOs to compress circuits for adiabatic evolution using the method from Ref. \cite{Mc_Keever_2023}. Ref. \cite{causer2023scalable} propose a similar tensor network optimization scheme to ours for quantum circuits using MPOs, providing technical details as well as benchmarks on different models. Our work is complementary as it combines both MPS and MPO with a focus on the performances on noisy quantum computers.}.

\begin{figure}[t]
\centering
\includegraphics[width = 1. \linewidth]{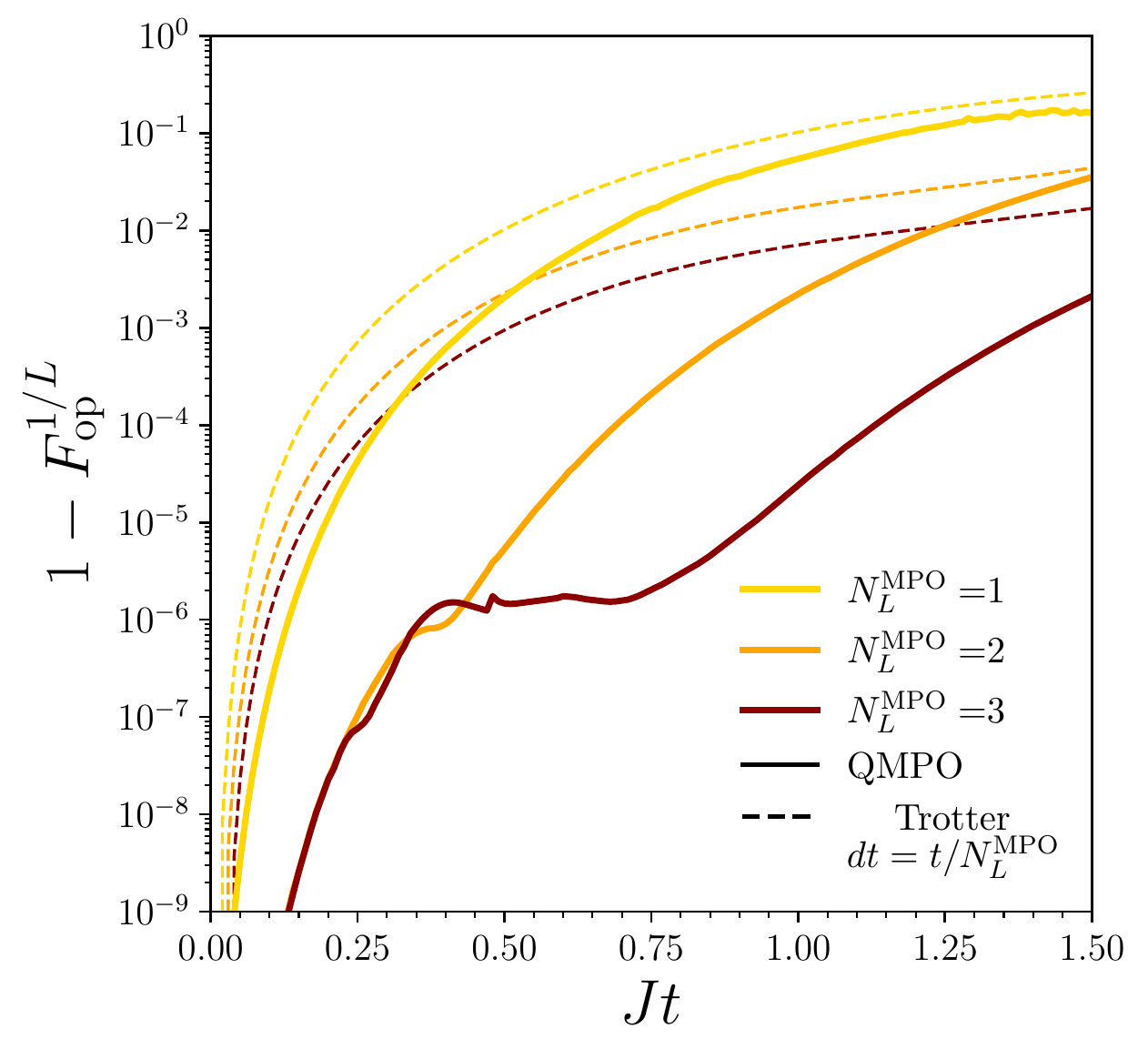}
\caption{The operator infidelity per site for the QMPO solutions for different number of layers $N_L^\text{MPO}$ against time for $L = 12$ sites. We compare the QMPO circuits (full line) with 1st order Trotter circuits of same depth (and therefore same number of quantum gates) with a corresponding time step $dt = t/N_L^\text{MPO}$ (dashed line).}
\label{fig:Fop_L12}
\end{figure}

In Fig. \ref{fig:Fop_L12}, we show the operator infidelity per site for a chain of $L=12$ sites for $N_L^\text{MPO} = 1$ and $2$ as a function of time $t  = ndt$, where $dt = 0.01$.  We perform optimization sweeps with a maximum number of sweeps set to $n_\text{sweep} = 100$. The convergence is reached when performing an optimization sweep does not improve the fidelity more than $10^{-8}$. We observe as expected that the short-time evolution is efficiently captured by the QMPOs, but degrades over time as the fixed number of layers bounds the entanglement entropy that is carried by the quantum circuits. We compare the performances of the QMPOs with 1st-order Trotter circuits employing the same number of layers (i.e. same number of gates) with a time step $dt = t/N_L^\text{MPO}$. While the QMPO errors are coming from the optimization process and the choice of the ansatz, the Trotter circuits suffer from errors inherent to the Trotter approximation. However, as shown in Fig. \ref{fig:Fop_L12}, the optimized QMPOs offer better operator fidelities than the Trotter circuits at a constant number of gates. Note that higher-order of the Trotter approximation can be used to lower the errors. While using higher-order Trotter circuits only increases the overhead of the QMPO optimization procedure, it results in the use of more quantum resources and therefore more noise when run on an actual device.

\section{Combining MPS/MPO and Noisy Quantum Computers}\label{sec:results}
In Section \ref{sec:qmps}, we displayed the capacity of QMPS to efficiently represent tensor network states as quantum circuits. This opens the possibility to extend classical simulations toward regimes where high-level of entanglement are required. By implementing QMPSs on quantum computers, it becomes possible to manipulate these states and generate more entanglement.  With the example of quantum dynamics, this offers the possibility to carry a simulation further in time by breaking the classical entanglement barrier.

In this Section, we propose a way to combine tensor network techniques with noisy quantum computers. We use the best of both worlds: short-time dynamics is efficently performed by MPSs, compiled into short-depth circuits, and then are propagated toward longer simulation time on a quantum device thanks to MPO-compressed circuits. In this framework, tensor networks assist the quantum simulation by providing all the ingredients (i.e. the initial state and the quantum circuits) to perform time evolution with fewer quantum resources. Again, this hybrid classical-quantum procedure is summarized in Fig. \ref{fig:schemeQMPSO}.

\subsection{Noisy QMPSO States}

To set our quantum simulations using QMPS and QMPO, that will call "QMPSO" simulations, we define a maximum simulation for the QMPS as $t_\text{max}^\text{MPS}$ as well as for the QMPO $t_\text{max}^\text{MPO}$. To reach a simulation time $t>t_\text{max}^\text{MPS}$, we decompose $t$ as $t = t_\text{max}^\text{MPS} + M\times t_\text{max}^\text{MPO} + \Delta t$, where $M = \lfloor \frac{t-t_\text{max}^\text{MPS}}{t_\text{max}^\text{MPO}} \rfloor$.  By denoting quantum circuits for the QMPS and QMPO by $U_\text{QMPS}(t)$ and $U_\text{QMPO}(t)$, the time-evolved state is given by

\begin{multline}
	\ket{\Psi_\text{QMPSO}(t)} = \\ U_\text{QMPO}(\Delta t) U_\text{QMPO}(t_\text{max}^\text{MPO}) ^M U_\text{ QMPS}(t_\text{max}^\text{MPS})\ket{\Psi_0}.
\end{multline}

In order to model noisy quantum computers, we employ a global depolarizing noise model. This phenomenogical description defines the density matrix of the noisy state $\rho$ as a mixture of the noiseless pure state density matrix $\rho_0 = |\Psi\rangle\langle\Psi|$ and a maximally mixed state $\mathds{1}/2^L$ as
\begin{equation}\label{eq:noise_model}
	\rho = \alpha \rho_0 + (1-\alpha)\frac{\mathds{1}}{2^L}, \text{   $\alpha \in [0,1]$},
\end{equation}
which is an approximation of a noisy state under a local depolarizing noise applied after each 2-qubit gate. We use $\alpha = e^{-\epsilon N_g}$, where $\epsilon$ is the two-qubit error rate and $N_g$ the number of gates used in the circuit \cite{Boixo_2018, Urbanek_2021, dalzell2021random}. This transformation models the gate errors with an incoherent stochastic noise. Although it does not capture all noise phenomena happening in a real quantum device, it describes one of the most prominent error sources of NISQ devices. Moreover, coherent noise can be converted into depolarizing noise with techniques like randomized compiling \cite{PhysRevA.94.052325}. Therefore simulating noisy quantum computers under such a model gives interesting and physically relevant insights into their performances.

\subsection{Fidelity}\label{subsec:fid}
To evaluate the quality of the states $\ket{\Psi(t)}$ or $\rho(t)$ obtained from the MPS techniques and noisy quantum computers, we calculate their overlap with a reference solution $\ket{\Psi_\text{ref}(t)}$. As in Sec. \ref{sec:qmps}, we choose this reference solution to be an exact solution time-evolved with a small time step $dt = 0.01$. For a noisy state $\rho(t)$, the fidelity is calculated as $F = \langle  \Psi_\text{ref}(t)|\rho(t)|\Psi_\text{ref}(t)\rangle \simeq e^{-\epsilon N_g} |\langle \Psi_\text{ref}(t)| \Psi(t)\rangle|^2$, according to Eq. (\ref{eq:noise_model}).

\begin{figure}[h]
	\centering
	\includegraphics[width=1.\linewidth]{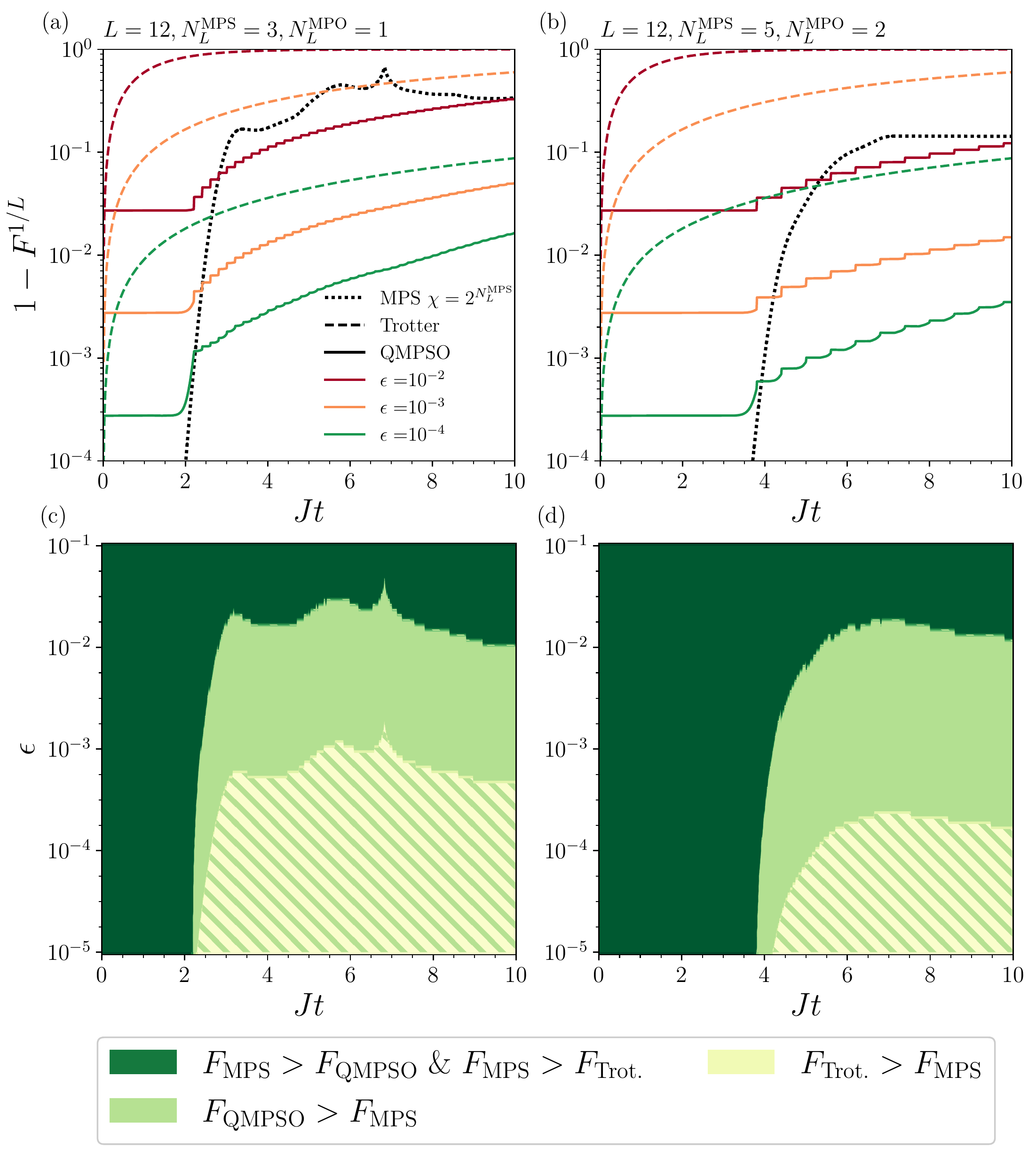}
	\caption{The infidelity per site for the QMPSO scheme for different error rate against the MPS solution for (a) $\chi_\text{MPS} = 2^3, N_L^\text{MPS} =3$, $t_\text{max}^\text{MPS} = 2.2$, $N_L^\text{MPO} =1$ and $t_\text{max}^\text{MPO} = 0.2$  and (b) $\chi_\text{MPS} = 2^5, N_L^\text{MPS} =5$,$t_\text{max}^\text{MPS} = 3.8$,  $N_L^\text{MPO} =2$ and $t_\text{max}^\text{MPO} = 0.6$. The dotted lines correspond to the MPS simulations, the dashed and full lines correspond respectively to noisy Trotter and QMPSO simulations. The plots (c) and (c) represent the diagram of advantage as a function of time and error rate. The domains are defined as follows; dark green: no advantage over the MPS, green: advantage with the QMPSO circuits, and light green: advantage with the Trotter evolution on a noisy quantum computer.}
	\label{fig:fidelityQMPSO}
\end{figure}

We compare the fidelities of noisy QMPSO states with noisy Trotter circuits and MPS simulations with Fig. \ref{fig:fidelityQMPSO}. In order to compare systems of different sizes, we again use the infidelity per site defined as $1 - F^{1/L}$ (see Supplementary Materials \ref{sec:siA} for system size comparison). We focus on a $L=12$ spin system, with a MPS bond dimension $\chi_\text{MPS} = 2^3$ ($2^5$), which corresponds to $N_L^\text{MPS} = 3/ N_L^\text{MPO} = 1$ ($N_L^\text{MPS} = 5/N_L^\text{MPO} = 2$) with Fig. \ref{fig:fidelityQMPSO}a (b), with a two-qubit error rate equal to $10^{-2},10^{-3}$, and $10^{-4}$ and a time step $dt$ set to $0.01$. First, we observe that a MPS provides always much better results for short-time dynamics, where they provide quasi-exact solutions. Second, using a QMPS as a starting point leads to better fidelities compared to pure Trotter quantum circuits, especially for short-time simulation. This is naturally expected as they use limited quantum resources while capturing efficiently the time evolution in this regime. Therefore, the main advantage of QMPSO with noisy quantum computers exists for a longer simulation time. The QMPO circuits allow to propagate the QMPS state further in time with much shorter circuits, leading to fidelities orders of magnitude higher compared to Trotter circuits. It is worth pointing out that in these results, the time step $dt$ has been set to $0.01$, which leads to large circuit depth but reduce significantly the Trotter errors. Finally, dedicating more classical resources, expressed here in terms of bond dimension, allows to manipulate more entanglement and therefore to provide better compressed quantum circuits, as illustrated with Fig.\ref{fig:fidelityQMPSO}a ($\chi_\text{MPS}= 2^3$) and b ($\chi_\text{MPS}= 2^5$).

Fig. \ref{fig:fidelityQMPSO}c\&d shows diagrams of the advantage of noisy quantum simulations over MPS solutions with the two maximum bond dimensions with respect to the simulation time and the noise level. We use the fidelity as a criterion for practical advantage, allowing us to compare errors coming for noisy hardware with truncation errors from MPS. It leads to the following conclusions: first, MPS naturally provide better simulation for short times, but also at longer times when the hardware errors are larger than the truncation errors. Therefore, there only exists a quantum advantage when targetting long simulation with sufficently low error rates. Trotter circuits necessitate exceptionally low error rates, especially when compared with MPSs of increasing bond dimension. The use of QMPSO circuits make possible to bridge the domains of MPS and quantum Trotter evolution, by beating MPS simulations with strongly reduced noise requirements.

We emphasize again that the effective noise map used here only takes into account gate errors. Despite being predominant in current devices, it does not capture all the relevant sources of error. In practice, complex phenomena such as crosstalk noise are also relevant noises occurring in quantum devices and play a major role in their performances \cite{Sarovar_2020,Georgopoulos_2021,perrin2023mitigating}. Reducing the circuit depth becomes even more crucial, and therefore using classical knowledge in the form of QMPSO circuits provides a promising pathway to use quantum devices.

\section{Experimental results}\label{sec:exp}
In the previous Section, we have considered quantum circuits with a depolarizing noise model. In this Section, we explore the benefits of combining tensor networks and quantum computers on actual quantum devices from IBM Quantum \cite{IBM}. 

To evaluate the performance of the experimental quantum simulations, we follow the dynamics of the transverse field Ising model by measuring the magnetization $\langle Z_i \rangle$ of each spin-1/2 in time and compare with both exact and truncated MPS solutions. Starting from an antiferromagnetic product state of size $L = 10$ we perform quantum simulations implementing Trotter and QMPSO circuits. We use optimized QMPSO circuits with $N_L^\text{MPS}= 3$ and $N_L^\text{MPO}= 1$  implemented as initial states, which encode time-evolved MPS states brought to their entanglement saturation time, here $t_{\text{max}}^{\text{MPS}}$ = 2.2  and $t_{\text{max}}^{\text{MPO}}$ = 0.5. The corresponding QMPS allows to capture at most 2/3 of the maximum entanglement entropy with respect to the system size. The time step of the Trotter steps $dt$ was set to $0.1$. 

To make the best use of the current quantum devices, we employ several error mitigation strategies, namely dynamical decoupling and pulse-scaling as implemented in qiskit \cite{Qiskit} and qiskit-research \cite{the_qiskit_research_developers_and_contr_2023_7776174}, which have been utilized in previous works \cite{ferris2022quantum,Kim_2023,PhysRevResearch.4.043027}. To efficiently implement the QMPSO circuits defined by optimized unitary matrices, we use the pulse-efficient decomposition as suggested in \cite{PhysRevResearch.3.043088}, which leverages native cross-resonance gates from IBM superconducting devices. Instead of using a CNOT-based decomposition, we express our two-qubit gates as a sequence of $R_{XX}$, $R_{YY}$, and $R_{ZZ}$ gates (and single qubit rotations) thanks to KAK decompositions \cite{tucci2005introduction}. Each of these two-qubit rotations is performed by a native cross-resonance pulse. A native cross-resonance pulse realizes a $R_{ZX}(\theta)$ gate in a shorter duration than a CNOT and therefore makes better use of the limited coherence time. To implement the Trotter circuits, we simplify the circuits by using only $R_{XX}$ and $R_{Z}$ gates. Similarly, the $R_{XX}$ gates are also performed by a pulse-efficient $R_{ZX}$ gate. 

\begin{figure*}
	\centering
	\includegraphics[width=1.\linewidth]{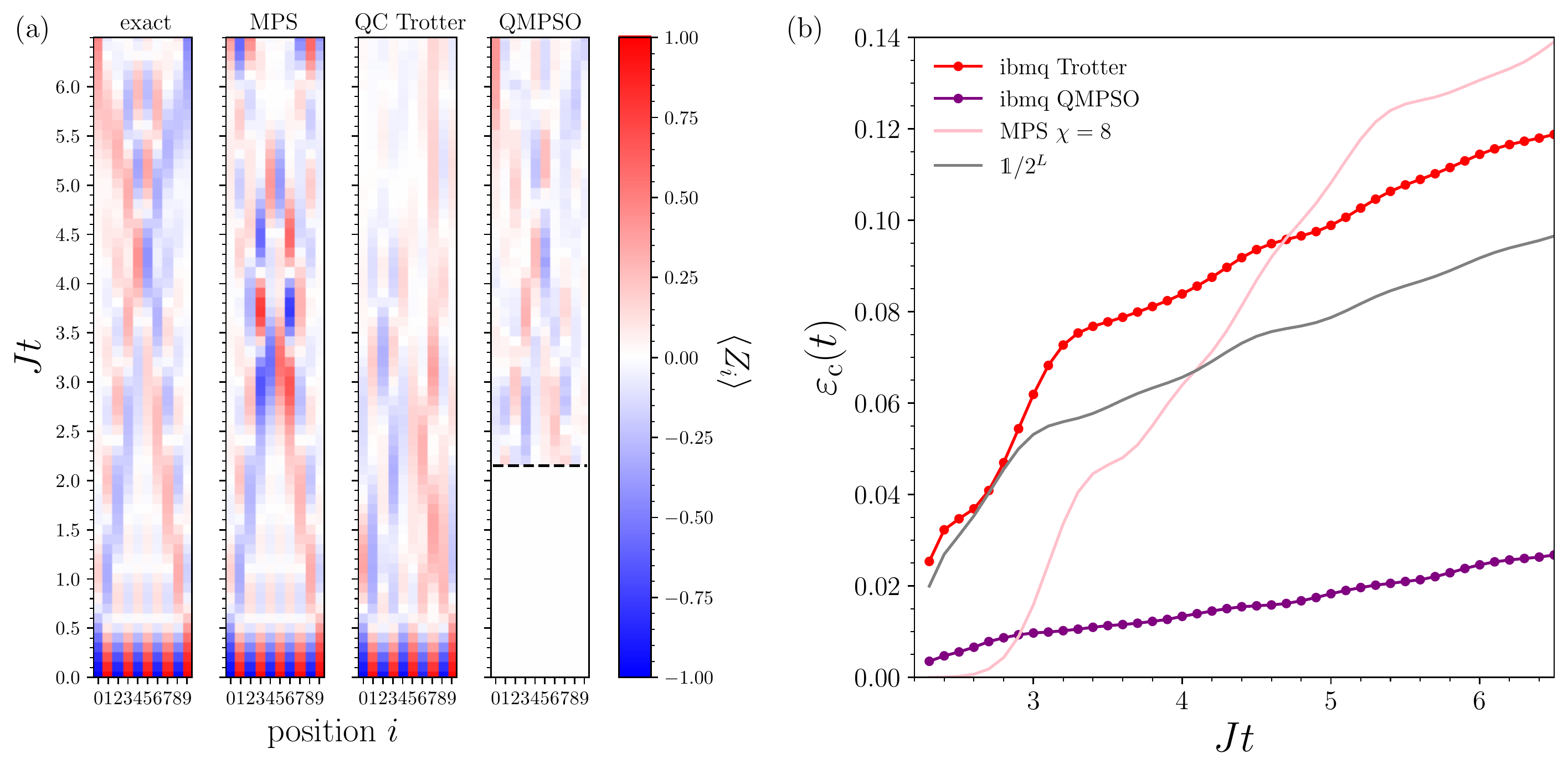}
	\caption{Evolution in time of the local magnetization in time for a $L=10$ quantum Ising chain. The QMPS is described with 3 sequential layers of two qubit gates, optimized via a MPS of dimension 8. (a) Colorplot for exact and truncated MPS solutions as well as Trotter  (``QC Trotter") and QMPSO simulations run on \texttt{ibmq\_kolkata}. (b) The cumulated error in local magnetizations over time for the MPS solution, the QPU results and a maximally mixed state $\mathds{1}/2^L$.}
	\label{fig:sz_exp}
\end{figure*}

Fig. \ref{fig:sz_exp} shows the results obtained from $\texttt{ibmq\_kolkata}$ for $L = 10$. We compute the local magnetization from the QPU and compare with both exact and MPS of bond dimension $\chi_\text{MPS} = 2^{N_L^\text{MPS}}$. In order to quantify the performances of the different methods, we use the cumulated error of the local magnetizations over time as in \cite{PhysRevResearch.4.043027}, that we define as:

\begin{equation}
	\varepsilon_c(t) = \frac{1}{t - t_{\text{max}}^{\text{MPS}}} \int_{t_{\text{max}}^{\text{MPS}}}^{t} d\tau
	\frac{1}{L} \sum_{i=1}^L |\langle Z_i\rangle(\tau) - \langle Z_i\rangle_{\text{exact}}(\tau)|^2
\end{equation}

We first observe that the time evolution with Trotter circuits allows us to recover the short-time dynamics, but, as expected, the quality of the results degrades for longer times as the circuit gets deeper. On the other hand, starting the simulation from a QMPS followed by QMPOs produce better results and beat the corresponding MPS that fails beyond $t_\text{max}^\text{MPS}$, especially for local magnetizations in the bulk of the spin chain where the entanglement entropy gets the largest. In Fig \ref{fig:sz_exp}a, we observe qualitatively the key features of the local magnetization for longer times than both Trotter evolution from the quantum device and from the MPS solution.  With Fig. \ref{fig:sz_exp}b, the cumulated error $\varepsilon_c(t)$ allows for more quantitative comparisons. The use of optimized QMPSO circuits translates into a significant improvement in the cumulated error. Finally, we compare the cumulated error with a maximally mixed state $\mathds{1}/2^L$, typically obtained from a deep noisy quantum circuit and containing no information. In our example, this corresponds to $\langle Z_i \rangle = 0$. This allows us to understand whether the expectation values obtained from the QMPS or Trotter carry relevant information or are dominated by noise. While Trotter simulations carry errors even higher than a maximally mixed state beyond $t_\text{max}^\text{MPS}$, the QMPSO simulations clearly provide meaningful results despite the presence of noise. The local magnetizations as a function of time are displayed in the Supplementary Materials \ref{sec:si_local_z_qpu}.

In this small instance, we demonstrate that a low-bond dimension MPS can be extended by a quantum computer, which allows one to study the dynamics of a quantum system beyond the MPS capabilities. Extending these experiments for larger system sizes combined with MPS of larger bond dimensions could lead to interesting insights into the capacities of current quantum devices.

\section{Conclusion}\label{sec:conclu}
In this manuscript, we have investigated the interplay between classical MPS techniques and noisy quantum computers for digital simulations of 1D spin systems. While MPSs are constrained by their bond dimension, which limits their capacity to represent highly entangled states, quantum computers suffer from experimental noise. Combining classically tractable MPSs and MPOs with efficient quantum circuits is therefore essential to get the most out of quantum devices. In order to bridge the gap between MPS and quantum simulations, we have used tensor network techniques to encode MPSs into classically optimized quantum circuits (QMPS) as well as compressing Trotter circuits with MPOs into shorter-depth circuits (QMPOs). We propose here to take QMPSs brought to their maximal capacities and use noisy quantum computers to break the MPS entanglement barrier. This relay from classical to quantum computers enables to make efficient use of limited quantum resources to reach a higher level of entanglement. To illustrate this protocol, that we call QMPSO, we have studied a global quench of the critical transverse field Ising Hamiltonian. In this system, the entanglement entropy exhibits a ballistic growth with time, which makes almost impossible the simulation of the system's dynamics over long timescales with MPS-based approaches. 

We have simulated noisy quantum circuits under a depolarizing noise and characterized the resulting states with their fidelity. We have compared the ability of quantum computers to beat MPS techniques with respect to target simulation times and the noise level of the devices. The use of QMPSO circuits can provide a significant improvement over the Trotter-Suzuki time-evolution as it reduces the circuit depth and reduces the error rate requirements for practical advantage. Finally, we tested experimentally this protocol on actual quantum devices from IBM Quantum \cite{IBM}. We observe improved results thanks to the use of QMPSO circuits compared to Trotter simulations. We also demonstrated in a small instance the ability of current devices to beat a low-bond-dimension MPS solution.  Comparing QPU computations with MPSs with different bond dimensions could also provide interesting insights into the performances of quantum computers with respect to a given classical complexity.

However, we emphasize that the MPS ansatz is particularly suited for 1D spin systems with local interactions, but becomes less efficient for 2D systems or systems with long-range interactions, while a quantum computer can be designed with the appropriate topology. Other tensor network topologies and their relationships with quantum circuits could also be considered.  Moreover, time-evolution with Trotter-Suziki circuits leads to high depth $D =  t/dt$ that are particularly inefficient in terms of noise and entanglement production. On the other hand, the QMPS representation shows that circuits of depth $D$ are sufficient to encode time-evolved states with an entanglement entropy $S_\text{vN} \lesssim D$. This gives hints on the optimal use of resources quantum algorithms should aim for. Variational quantum algorithms \cite{yuan_theory_2019} combined with QMPSs can offer interesting alternative solutions to study the quantum dynamics with appropriate circuit depths with respect to the required entanglement level.

\section*{Acknowledgment}
We are grateful to the Association Nationale de la Recherche et de la Technologie for funding. We acknowledge the use of IBM Quantum services for this work. The views expressed are those of the authors, and do not reflect the official policy or position of IBM or the IBM Quantum team.
\bibliography{bibliography}

\subfile{supplementary_materials}


\end{document}

%% file: supplementary_materials.tex
\section*{Supplementary Materials}

\subsection{QMPS and QMPO vs system size}\label{sec:siA}
Here we explore the performances of the  QMPS and QMPO optimization as a function of the system size. To do so, we optimize quantum circuits for different system sizes and number of layers. We run a TEBD algorithm on MPS and MPO with time step $dt = 0.01$ and optimize quantum circuits every $\Delta t = 0.5$ for QMPS and $0.25$ for QMPO. The maximum number of sweeps was set to 2000 for QMPS and 1000 for QMPO. 

\begin{figure}[h]
	\centering
	\includegraphics[width = 1.\linewidth]{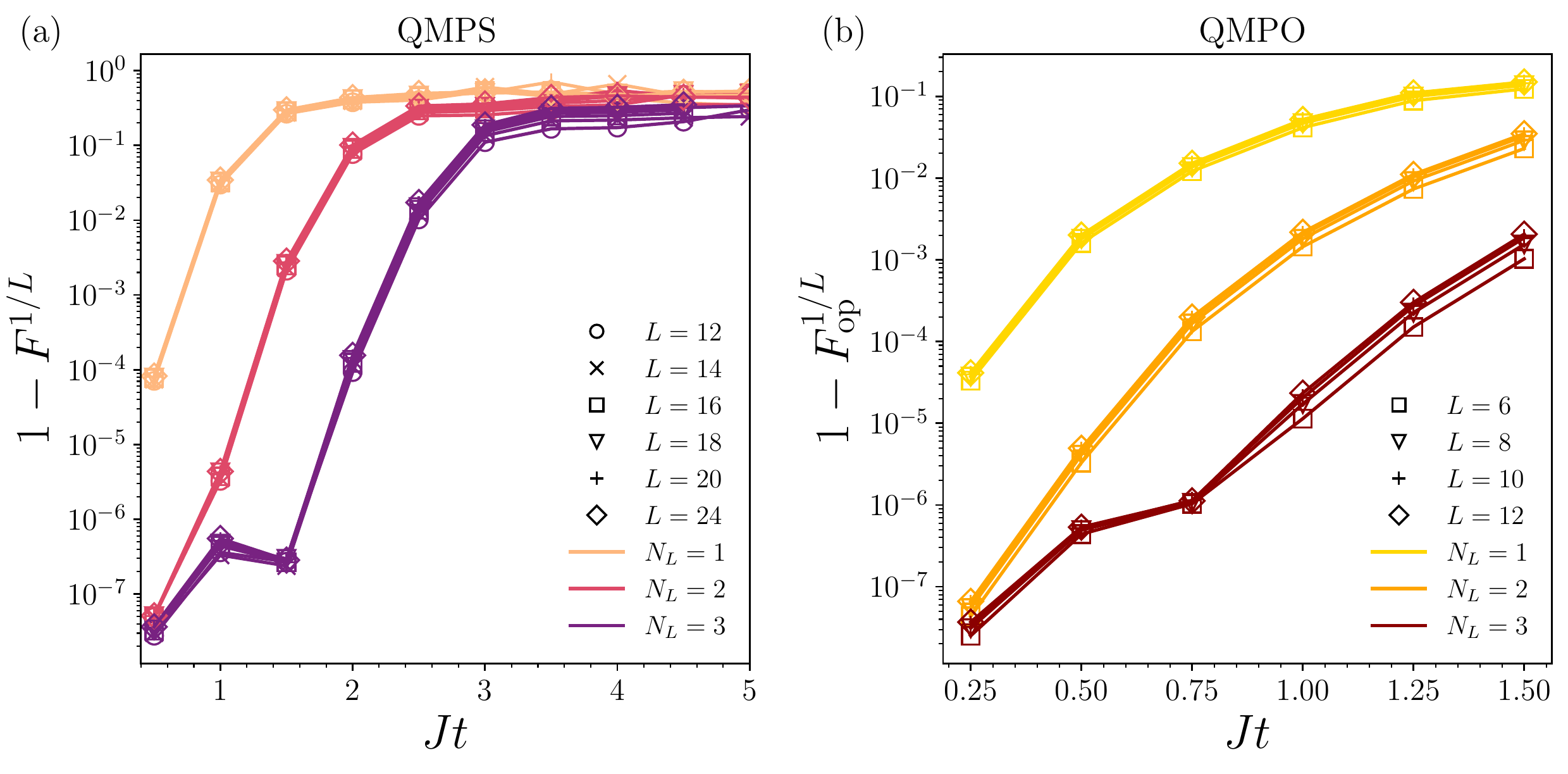}
	\caption{The infidelity per site for (a) QMPSs and (b) QMPOs optimized with different number of layers and system size.}
	\label{fig:fqmpso_vs_l}
\end{figure}

Fig. \ref{fig:fqmpso_vs_l} shows the infidelity per site obtained for the QMPS and QMPO for various system sizes. We observe no clear dependance with respect to the length $L$ of the system. This indicates that the key parameter in these simulations is the bond dimension which then set the number of layers of the circuits to optimize. This can be explained by the fact that the bond dimension acts locally by restricting the dimensions of the local tensors of a MPS or MPO.
Moreover, the noise model employed here also leads to a error per site invariant with the system size. Indeed, a depolarizing noise model typically decrease the noiseless fidelity by a factor $e^{-\epsilon N_g}$, where $\epsilon$ is the two-qubit gate error rate and $N_g$ the number of two-qubit gates. In our work, $N_g$ is equal to $(L-1)\times D$, with $D$ is the depth of the circuit, i.e. the number of QMPS/QMPO layers or Trotter steps. When looking at the error per site, this factor quickly becomes independent with $L$ as $(e^{-\epsilon D(L-1)})^{1/L} \simeq e^{-\epsilon D}$ for $L \gg 1$. Together with Fig. \ref{fig:fqmpso_vs_l}, this allows us to expect similar performances of the QMPSO circuits when targeting higher system sizes and bond dimension than the ones explored in this work.
 \subsection{Entanglement production}\label{sec:siB}
The key bottleneck of tensor network techniques lies in their limited capacity to encode entanglement, while quantum computers naturally use entanglement. However, the noise inherent to quantum devices prevents them from running deep quantum circuits, and therefore they are also limited in terms of entanglement production. After having looked at the fidelity, another interesting quantity to consider is the entanglement as a performance metric to compare MPS solutions with noisy quantum circuits.  
The depolarizing noise model used here captures the effect of noise on entanglement production. Indeed, it can be understood as a process of non-destructive local measurements, where the density gets mixed with states with a lower entanglement level. Therefore, after applying a two-qubit gate, there exists a competition between the entanglement generated by the gate and the noise occurring that reduces the quantum correlations. 
To quantify the growth of quantum correlations in a noisy circuit, we use the operator entanglement entropy of the density matrix which was used as a benchmark metric in \cite{noh_efficient_2020, zhang_entanglement_2022}, in the context of the simulation of noisy circuits with MPO techniques. In the case of density matrix calculations, the operator entanglement entropy $S_{\text{op}}$ can be obtained thanks to a Schmidt decomposition of the density matrix $\rho$ for a cut between subsystems $A$ and $B$:
\begin{equation}
	\rho = \sum_{\alpha=1}^{2^L} \lambda_\alpha \rho^A_\alpha \otimes \rho^B_\alpha,
\end{equation}
where $\rho_\alpha^A$ and $\rho_\alpha^B$ are reduced density matrices for the two subsystems of pure states, and $\{\lambda_\alpha\}$ are the Schmidt values. Here $\lambda_\alpha$ corresponds to the classical probability of finding the state in $\rho^A_\alpha \otimes \rho^B_\alpha$ in the mixed state $\rho$. From this decomposition, the operator entanglement entropy is defined as
\begin{equation}
	S_{\text{op}} = -\sum_\alpha \frac{\lambda_\alpha^2}{\sum_\beta \lambda_\beta^2} \log_2 \left(\frac{\lambda_\alpha^2}{\sum_\beta \lambda_\beta^2}\right).
\end{equation}
As detailed in \cite{noh_efficient_2020}, it is important to note that although $S_{\text{op}}$ does not distinguish between quantum and classical correlations, it vanishes for the maximally mixed state 
$\rho  = \mathds{1}/2^L$ that is typically obtained in the deep circuit limit under a depolarizing noise.

\begin{figure}[h]
	\centering
	\vspace{10pt}
	\includegraphics[width=1.\linewidth]{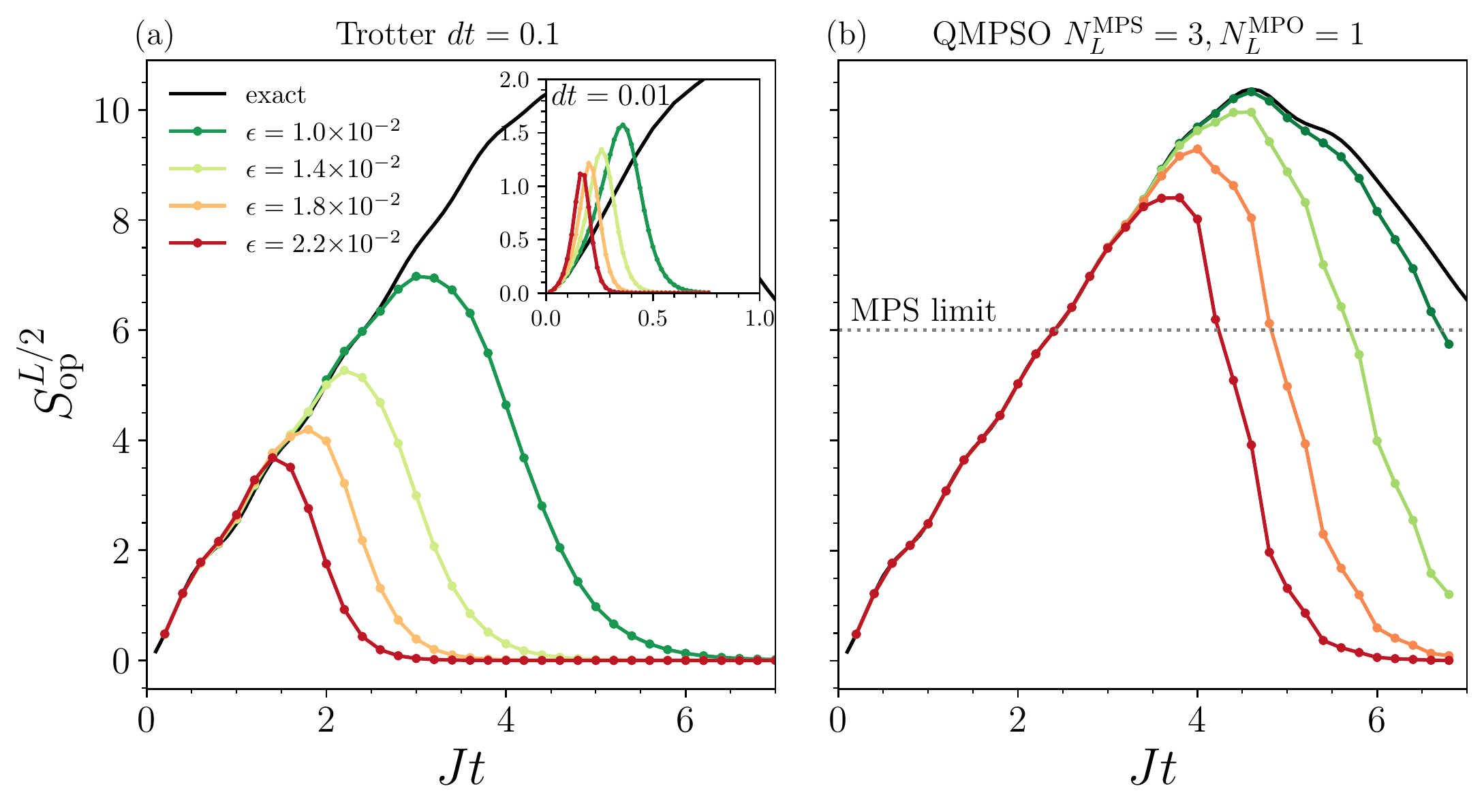}
	\caption{Operator entanglement entropy in time for noisy states with $L =12$ sites, (a) for Trotterized evolution with $dt =0.1$ ($dt =0.01$ in the inset plot), and (b) for QMPSO simulation with $N_L^\text{MPS} = 3, N_L^\text{MPS} = 1$. The dotted grey line correspond to the maximum entanglement for the corresponding MPS of bond dimension $\chi = 2^{N_L^\text{MPS}}$.}
	\label{fig:Sop}
\end{figure}

Taking a depolarizing noise model, we compute the operator entanglement entropy for Trotter and QMPSO simulation with Fig. \ref{fig:Sop} using realistic noise levels. We observe that the Trotter circuits are failing to prepare states with quantum correlations beyond what MPS capabilities (with respect to the bond dimension), meaning that such noisy states could also be efficiently simulable with density matrix simulations based on MPO \cite{noh_efficient_2020, zhang_entanglement_2022} for example. On the other hand, the noisy QMPSO circuits are capable of preparing entangled noisy states beyond tensor network reach, which also consists of an advantage over classical simulation.

\subsection{Experimental data}\label{sec:si_local_z_qpu}
\begin{figure*}
	\centering
	\includegraphics[width=1.\linewidth]{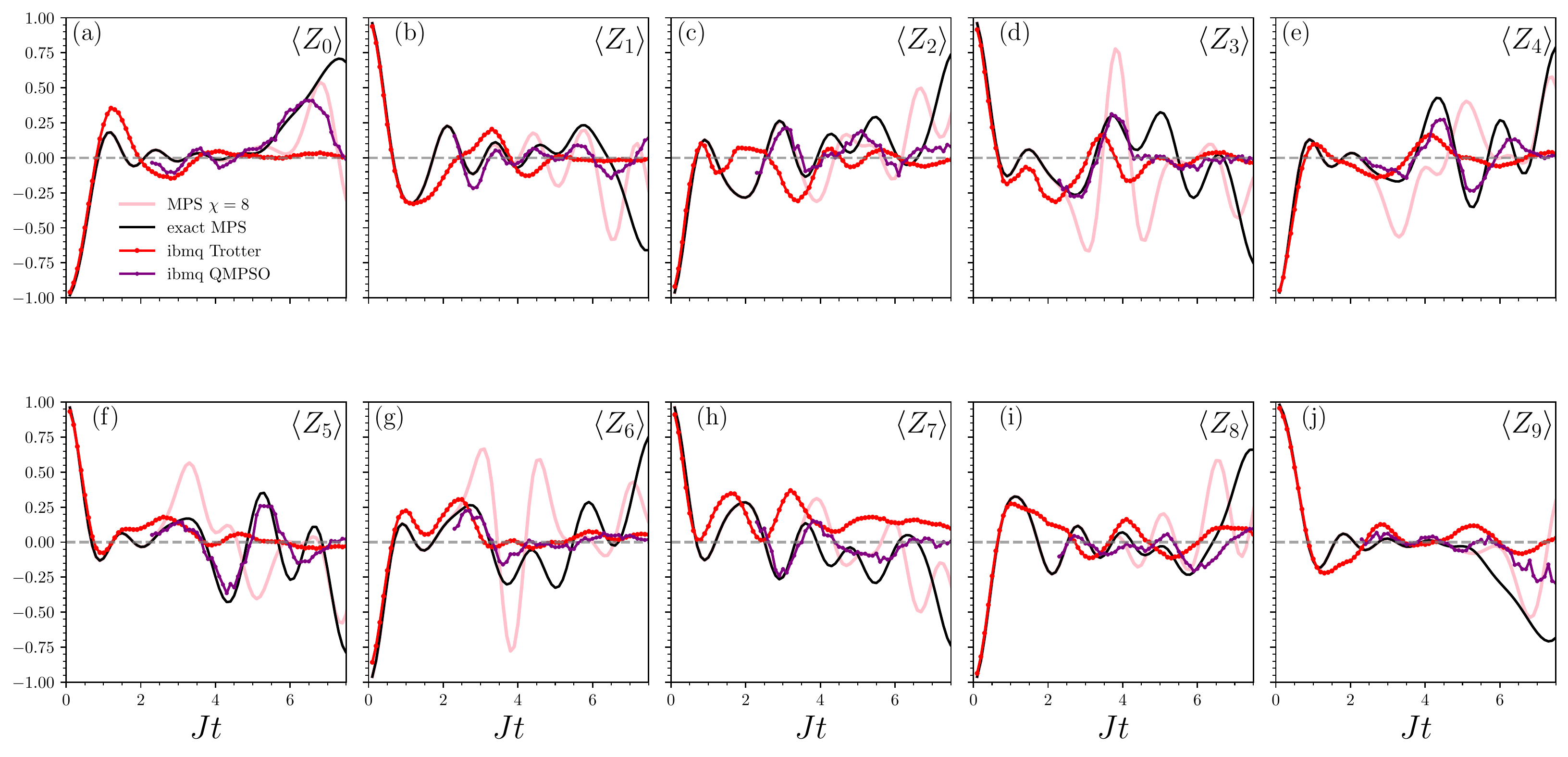}
	\caption{Local magnetization of each site from the 10-qubit simulation performed on $\texttt{ibmq\_kolkata}$. }
	\label{fig:sz_ibmq_individual}
\end{figure*}

Here we show the local magnetization of the quantum simulations carried on $\texttt{ibmq\_kolkata}$ for each site with Fig. \ref{fig:sz_ibmq_individual}. Trotter circuits capture the right dynamics only for short times, while the QMPSO simulation provides a better match with exact magnetizations at times immediately following time $t_{\text{max}}^{\text{MPS}}$ and capture the correct behavior over longer times. We also observe discrepancies with respect to sites that manifest the quality variability of the qubits and their couplings. The experiments were performed on the 23th (QMPSO) and 24th (Trotter) of October 2023.

